\documentclass[]{emulateapj}

\def\Mpc{\ {\rm Mpc}}

\def\kms{{\ }{\rm km}\,{\rm s}^{-1}}
\def\vc{$v_{\rm c}$}
\def\sigmav{\sigma_{\rm v}}
\def\Msun{M_{\odot}}
\def\Mstel{M_{\ast}}
\def\lcdm{$\Lambda$CDM}

\def\i{$i$-}

\def\resp{respectively}
\def\sersic{S\'ersic }
\def\to{$\to$\ }
\def\Y07{\citet{Yang07a}}
\def\HI{H{\footnotesize I}}
\def\zgrp{z_{\rm grp}}
\def\Mgrp{M^\ast_{\rm grp}}
\def\Mhalo{M_{\rm halo}}
\def\Nspec{N_{\rm spec}}
\mathchardef\mhyphen="2D
\def\Fetype{F_{\rm E\mhyphen type}}
\def\Fltype{F_{\rm L\mhyphen type}}
\def\Nbody{$N$-body}

\usepackage{MnSymbol}
\usepackage{wasysym}
\usepackage{latexsym}
\usepackage{amssymb}
\usepackage{graphicx}
\usepackage{mathrsfs}
\usepackage{amsmath}
\usepackage[bottom]{footmisc}
\usepackage{color}
\usepackage[breaklinks=true]{hyperref}
\shortauthors{ABRAMSON ET AL.}
\shorttitle{THE VELOCITY FUNCTION OF GROUP GALAXIES}

\begin{document}

\title{THE CIRCULAR VELOCITY FUNCTION OF GROUP GALAXIES}

\slugcomment{Submitted to The Astrophysical Journal}

\author{
Louis E. Abramson\altaffilmark{1,2,$\ast$}, 
Rik J. Williams\altaffilmark{2},
Andrew J. Benson\altaffilmark{2}, 
Juna A. Kollmeier\altaffilmark{2},
and John S. Mulchaey\altaffilmark{2}
}


\begin{abstract}

A robust prediction of \lcdm\ cosmology is the halo circular velocity function (CVF), a dynamical cousin of the halo mass function. The correspondence between theoretical and observed CVFs is uncertain, however: cluster galaxies are reported to exhibit a power-law CVF consistent with \Nbody\ simulations, but that of the field is distinctly Schechter-like, flattened compared to \lcdm\ expectations at circular velocities $v_{\rm c}\lesssim 200\kms$. Groups offer a powerful probe of the role environment plays in this discrepancy as they bridge the field and clusters.  Here, we construct the CVF for a large, mass- and multiplicity-complete sample of group galaxies from the Sloan Digital Sky Survey.  Using independent photometric \vc\ estimators, we find {\it no} transition from field- to $\Lambda$CDM-shaped CVF above $v_{\rm c} = 50 \kms$ as a function of group halo mass.  All groups with $12.4 \lesssim \log \Mhalo / \Msun \lesssim 15.1$ (Local Group analogs to rich clusters) display similar Schechter-like CVFs marginally {\it suppressed} at low-\vc\ compared to that of the field.  Conversely, some agreement with \Nbody\ results emerges for samples saturated with late-type galaxies, with {\it isolated} late-types displaying a CVF similar in shape to \lcdm\ predictions.  We conclude that the flattening of the low-\vc\ slope in groups is due to their depressed late-type fractions -- environment affecting the CVF only to the extent that it correlates with this quantity -- and that previous cluster analyses may suffer from interloper contamination. These results serve as useful benchmarks for cosmological simulations of galaxy formation.\\

\end{abstract}

\keywords{
cosmology: observational --
cosmology: simulations -- 
galaxies: groups -- 
galaxies: circular velocity function -- 
galaxies: mass function
}

\altaffiltext{1}{
Department of Astronomy \& Astrophysics and Kavli Institute for Cosmological Physics, The University of Chicago, 5640 South Ellis Avenue, Chicago, IL 60637, USA
}
\altaffiltext{2}{
The Observatories of the Carnegie Institution for Science, 813 Santa Barbara Street, Pasadena, CA 91101, USA
}
\altaffiltext{$\ast$}{
\href{mailto:labramson@uchicago.edu}{\tt labramson@uchicago.edu}.
}


\section{Introduction}

The establishment of the \lcdm\ cosmological paradigm over the past several decades represents an unprecedented step towards creating a unified, consistent description of the universe.  At super-galactic scales this theory has withstood repeated observational tests and its parameters are now tightly constrained \citep[e.g.,][]{Riess11, Freedman12, WMAP9, SPTCosmoParams13, PlanckParams13}.

However, $\Lambda$CDM-based models have not achieved similar success in explaining the observed properties of galaxies.  For example, cosmological simulations of galaxy formation have not yet proven their ability to generate stellar mass or luminosity functions -- perhaps the most rudimentary encapsulations of these properties -- which closely resemble observations (see e.g., \citealt{Benson03, Oppenheimer10, Dave11}, but cf. \citealt{Vogelsberger14}).  That said, because the astrophysical processes that define observations are difficult to capture numerically, it is unclear whether such discrepancies point to serious failings of the \lcdm\ picture or merely current modeling techniques.

To progress, meeting-points for theory and observation that are less sensitive to baryonic physics -- and therefore uncertainties in, e.g., feedback processes -- would be useful. Because circular velocity, \vc, is in principle both a shared observable and agnostic to the form of the gravitating matter, the galaxy circular velocity function, CVF or $\phi(v_{\rm c})$, provides appealing common ground.  

\begin{deluxetable*}{cccc}
\tablecolumns{4}
\tablecaption{Group and Galaxy Parameters}
\tablehead{
\colhead{{ Quantity}} &
\colhead{Unit} &
\colhead{Source\tablenotemark{a}} &
\colhead{Definition / Comment}
}
\startdata
$r$				& AB mag		& 2 & Petrosian aperture\\
$M_r$			& AB mag		& 2 & Extinction- and $k$-corrected to $z = 0.1$; also $^{0.1}M_r$\\
$g-r$				& AB mag		& 2 & Extinction- and $k$-corrected to $z = 0.1$; also $^{0.1}(g - r)$\\
$M_{\ast}$		& $\Msun$	& 2 & Galaxy stellar mass\\
$z$				& $\cdots$		& 1 & Galaxy redshift\\
$\zgrp$			& $\cdots$		& 1 & Group luminosity-weighted $\langle z \rangle$\\
$\Mgrp$			& $\Msun$	& 1 & Group stellar mass proxy (adjusted to $h = 1$)$\rm{^b}$\\
$\Mhalo$			& $\Msun$	& 1 & Luminosity-ranked group halo mass (adjusted to $h = 1$)$\rm{^c}$\\
MMG				& $\cdots$		& 1 & Most-massive group galaxy (binary flag)\\
$n$				& $\cdots$		& 2 & $r$-band \sersic index\\
$R_0$			& arcsec		& 2 & $r$-band half-light radius (from \sersic fit)\\
$b/a$			& $\cdots$		& 2 & $r$-band axis ratio (exponential fit)\\
$\sigmav$			& $\kms$		& 2 & Galaxy spectroscopic stellar velocity dispersion\\
$\Nspec$			& $\cdots$		& 3 & \textnumero\ spectroscopic members per group; ``richness"\\
$F_{\rm E/L\mbox{-}type}$	& $\cdots$		& 3 & Early-/late-type fraction {\it within a group}
\enddata
\tablenotetext{a}{(1) Y07; (2) VAGC; (3) Derived}
\tablenotetext{b}{Completeness-corrected total stellar mass of galaxies with $M_r \leq -19.5$ (see Equation 13 of Y07).}
\tablenotetext{c}{Based on abundance matching.}
\label{tbl:params}
\end{deluxetable*}

Currently, the correspondence between observed and theoretical CVFs is unclear.  Dark matter simulations predict that the (halo) CVF of the general field should be described by a power-law with slope $\alpha \sim -3$ to $-4$ \citep[][adapted from \citealt{Kravtsov04,Zavala09}]{Klypin99, Moore99b, Blanton08}. Yet, observations of field galaxies show the CVF (or the related velocity dispersion function) to be Schechter-like in form \citep{PressSchechter74, Schechter76} with a substantially flatter slope at $v_{\rm c} \lesssim 200 \kms$ \citep[][but cf. \citealt{Blanton08}]{Gonzalez00, Sheth03, Choi07, Chae10, Zwaan10, Bezanson11}.

Compounding matters, although galaxies in the field seem not to conform to theoretical expectations, at least one observation suggests that those in clusters {\it do}.  Using photometric \vc\ estimators, \citet[][hereafter D04]{Desai04} found { cluster galaxies} to exhibit a power-law CVF with $\alpha = -2.4 \pm 0.8$, consistent with model predictions.

If cluster, but not field galaxies have a CVF that is { apparently} well-described by \lcdm, the natural question to ask is: Where does the break-down occur?  To address this question we can turn to groups of galaxies.

Previously, \citet{Pisano11} examined the group galaxy CVF for a small sample of Local Group analogs using \HI\ data.  These authors also found the low-\vc\ slope to be flattened relative to \lcdm\ predictions (i.e., field-like).  However, although their investigation reached $v_{\rm c} \lesssim 10 \kms$ ($M_{\rm \HI} \sim 7 \times 10^5 \Msun$) it covered only 61 galaxies in 6 groups confined to a thin slice of group demographics: loose associations with mass $\log\Mhalo/\Msun \lesssim 13.6$, $N_{\rm gals} \lesssim 20$, and high late-type fractions.  Since groups span much broader ranges in mass, richness, and late-type fraction than this \citep[e.g.,][]{ZabludoffMulchaey98, Weinmann06}, each of which might drive the CVF towards or further away from that reported for clusters or \lcdm, a broader study is warranted.  

Here, we take advantage of the extensive Sloan Digital Sky Survey \citep[SDSS --][]{York00} to probe the CVF in a large set of such diverse environments.

We proceed as follows: Section 2 describes the group/galaxy catalogs and various issues associated with their use, Section 3 outlines the estimation of \vc, Section 4 presents our results, Section 5 our discussion, and Section 6 our conclusions. All magnitudes are quoted in the SDSS Petrosian system. All fits were computed using {\tt MPFIT} in {\tt IDL} \citep{MPFIT}.


\section{Data}

We base our investigation on a mass- and richness-complete sample of groups from the catalog of \citet[][hereafter Y07]{Yang07a}.  This is a well-established catalog \citep[see e.g.,][]{vandenBosch08, Moster10, WetzelWhite10, PengLilly12} but we have verified that our main results are robust to catalog selection by repeating our analysis using that of \citet[][see Section 4.4 below]{Tinker11}.  We do not adopt this as our primary sample because it is limited to galaxy stellar masses $\log\Mstel/\Msun \geq 9.4$ and thus provides less leverage on the CVF at low \vc.  The group/galaxy catalogs and our sample inclusion criteria are described below.

\subsection{The Group Catalog}
\label{sec:grpcat}

The Y07 group catalog is derived from the spectroscopic sample from the fourth SDSS data release \citep[DR4 --][]{SDSS_DR4}.  We use ``Sample II", which incorporates 7091 galaxies from other sources as compiled in the NYU Value Added Galaxy Catalog \citep[VAGC --][]{Blanton05VAGC, PadVAGC08}.  The sample comprises 369,447 galaxies spectroscopically assigned to 301,237 groups at $0.01 \leq z \leq 0.20$.

Although ``Sample III" contains photometric near-neighbors and is thus, in principle, deeper and more complete, the added sources suffer foreground/background contamination rates near $40\%$. We avoid this sample as the presence of interlopers might mask significant environmental effects.

Group properties (see Table \ref{tbl:params}) are quoted from the Y07 catalog and  adjusted to VAGC cosmology where appropriate (see Section \ref{sec:galcat}). We refer the interested reader to Y07 for details regarding group identification and characterization, but note a few key items here.

Group stellar masses, $\Mgrp$, reflect the sum of galaxy stellar masses, $\Mstel$, for members with $M_r \leq -19.5$.  Halo masses, $\Mhalo$, are assigned via abundance matching to mock catalogs using either $\Mgrp$ or a similarly determined characteristic luminosity.  We quote { the luminosity-derived $\Mhalo$}, but base our analysis entirely on $\Mgrp$, so the choice is largely superficial.  Scatter between $\Mhalo$ estimates is $\lesssim 0.1$ dex.

A large number of groups contain no galaxies with $M_r \leq -19.5$ and thus lack stellar or halo mass estimates. Simply excluding such ``massless groups" has no effect on MAIN results, but does significantly alter those from the ALL sample (see next section and Appendix \ref{sec:AB}).  We have therefore assigned these systems -- biased by construction towards isolated galaxies with $\log\Mhalo/\Msun \lesssim 11.9$ -- masses based on a fit to the relationship between total member $\Mstel$ and $\Mgrp$ or $\Mhalo$ for groups where the latter quantities are known.

Finally, Y07's selection process leads to the identification of many ``groups" harboring only one member ($\Nspec = 1$; see Table \ref{tbl:params}).  As the statistics above show, such systems constitute the vast majority of the catalog.  However, we generally use ``group" to mean systems with richness $\Nspec \geq 2$.

\subsection{The Galaxy Catalog}
\label{sec:galcat}

Subsequent to Y07, the VAGC has been updated.  Galaxy properties are thus obtained by associating Y07 sources to the DR7 \citep{SDSS_DR7} VAGC using M.\ Blanton's {\tt SPHEREMATCH}, part of the {\tt IDLUTILS} library.\footnote[3]{\url{www.sdss3.org/dr8/software/idlutils.php}.}  All sources match to a counterpart within $2\farcs0$.

{\it All galaxy properties are quoted directly from the VAGC} (see Table \ref{tbl:params}), and so reflect $h \equiv H_0/100 \kms \Mpc^{-1} = 1$. Y07 take $h = 0.73$. As this affects only $\Mgrp$ and $\Mhalo$, we simply adjust these quantities to the VAGC cosmology, adopting $(H_0,\Omega_{\rm m}, \Omega_{\Lambda}) = (100 \kms \Mpc^{-1},0.27,0.73)$.

The full VAGC also serves as our field control sample.  We refer to this when discussing ``the field" below.

\subsection{Incompleteness}
\label{sec:incompleteness}

Understanding incompleteness is key to any CVF measurement.  As we are interested in the ``conditional" CVF of galaxies in groups of various characteristics, $\phi(v_{\rm c}\,|\, \Mgrp, \Nspec, \dots)$, both galaxy and group incompleteness affect our results. Unfortunately, each depends significantly on the other: flux limits ($r_{\rm SDSS}^{\rm lim} \approx 18$) and fiber collisions ($\Delta\theta_{\rm SDSS}^{\rm min} \approx 55\arcsec$) can remove galaxies, which can prevent group identification, which can remove more galaxies from the catalog. Appendices \ref{sec:AA} and \ref{sec:AF} present detailed discussions of our treatment of these issues; it is sufficient to note here that, because we are ultimately interested in galaxy-based quantities, galaxy incompleteness largely defines our sample.

The simplest, least model-dependent way to mitigate galaxy incompleteness is to limit analyses to redshifts where (1) the spectroscopic catalog is sufficiently \vc-complete, and (2) groups span solid angles large enough to ensure most members have no neighbors within the minimum SDSS fiber spacing of $55''$ \citep[][]{Zehavi02}. 

\begin{figure}[t!]
\hskip -0.4cm
\includegraphics[width = 1.15\columnwidth]{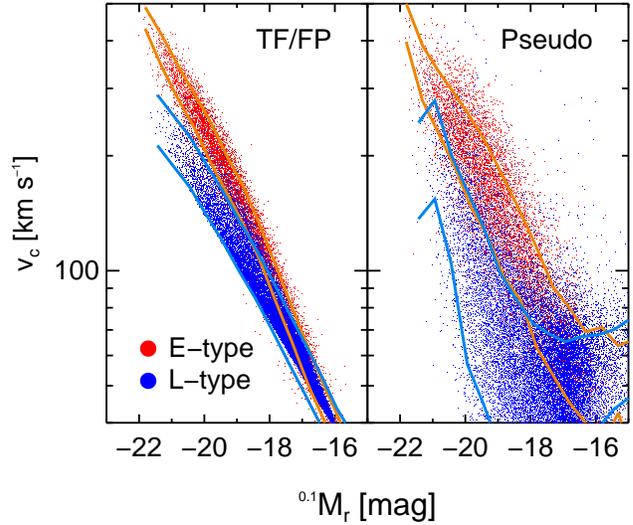}
\caption{Luminosity--\vc\ relationships for early- and late-type galaxies (see Section 3.1).  {\it Left}: Inverse Tully-Fisher (TF) and Fundamental Plane (FP) estimates.  {\it Right}: pseudo-dispersion estimates.  Solid lines show adopted $1\sigma$ uncertainties (see Section 3.3).  Absolute $r$-band magnitudes are $k$-corrected to $z = 0.1$.  See Appendix \ref{sec:AC} and Figure \ref{fig:metric_comp} for metric cross-comparisons.\vspace{0.25em}}
\label{fig:vc_err}
\end{figure}

We adopt a redshift cut-off of $\zgrp \leq 0.03$. Using our scaling relations (Section \ref{sec:TFFP}, Appendices \ref{sec:AC} and \ref{sec:AD}, Figure \ref{fig:vc_err}), the corresponding $M_r = -16.8$ luminosity limit translates to a \vc\ completeness limit of $v_{\rm c} = 50 \kms$ (roughly $\log \Mstel/\Msun = 8.7$); i.e., systems similar to the Small Magellanic Cloud \citep{Stanmirovic04}. The mean nearest-neighbor separation for galaxies in these groups is $\sim 200''$--$300''$, well beyond the fiber-collision limit. Measurements from galaxies in overlap regions suggest that the small fraction of sources excluded by collisions are unbiased in color and luminosity relative to the rest of the sample.

We have verified that our main conclusions hold using sub-samples truncated as high as $\log\Mstel/\Msun \geq 9.4$.

\subsection{The Samples}

We analyze three samples below:
\begin{enumerate}
	\item{FIELD -- Field control sample; all 31359 galaxies in the DR7 VAGC with $0.01 \leq z \leq 0.03$.}
	\item{ALL -- All Y07 systems in the same redshift interval; 15991 galaxies in 11724 ``groups" of $\Nspec \geq 1$.}
	\item{MAIN -- Main group sample; 2835 galaxies in 372 groups with $0.01 \leq \zgrp \leq 0.03$, $\log \Mgrp / \Msun \geq 11.0$, and $\Nspec \geq 2$.}
\end{enumerate}

MAIN groups span the halo mass range $12.4 \lesssim \log \Mhalo / \Msun \lesssim 14.3$ with $\langle \log\Mhalo/\Msun \rangle \simeq 12.9$.  To avoid additional scaling dependencies, however, we quote $\Mgrp$ almost exclusively.  A useful number for converting to $\Mhalo$ is the average $\log(\Mhalo / \Mgrp$), or 1.7.


\section{Circular Velocity Estimation}

Constructing the CVF requires estimating \vc.  The VAGC provides spectroscopic velocity dispersions, $\sigmav$, for most Y07 sources.  One could simply scale these  since $v_{\rm c} = \sqrt{2}\sigmav$ assuming (as we do in places) isothermal spherical halos \citep{BinneyTremaine}.  However, there are two important drawbacks to using the {\it spectroscopic} dispersions: (1) it prevents analyses below $\sigmav \approx 70 \kms$, where individual SDSS measurements become unreliable \citep[e.g.,][]{Eisenstein03}; (2) the mapping between measured $\sigmav$ and halo \vc\ is unclear for rotationally supported galaxies.

\begin{figure*}[t!]
\centering
\hskip -0.75cm
\includegraphics[width = .95\linewidth]{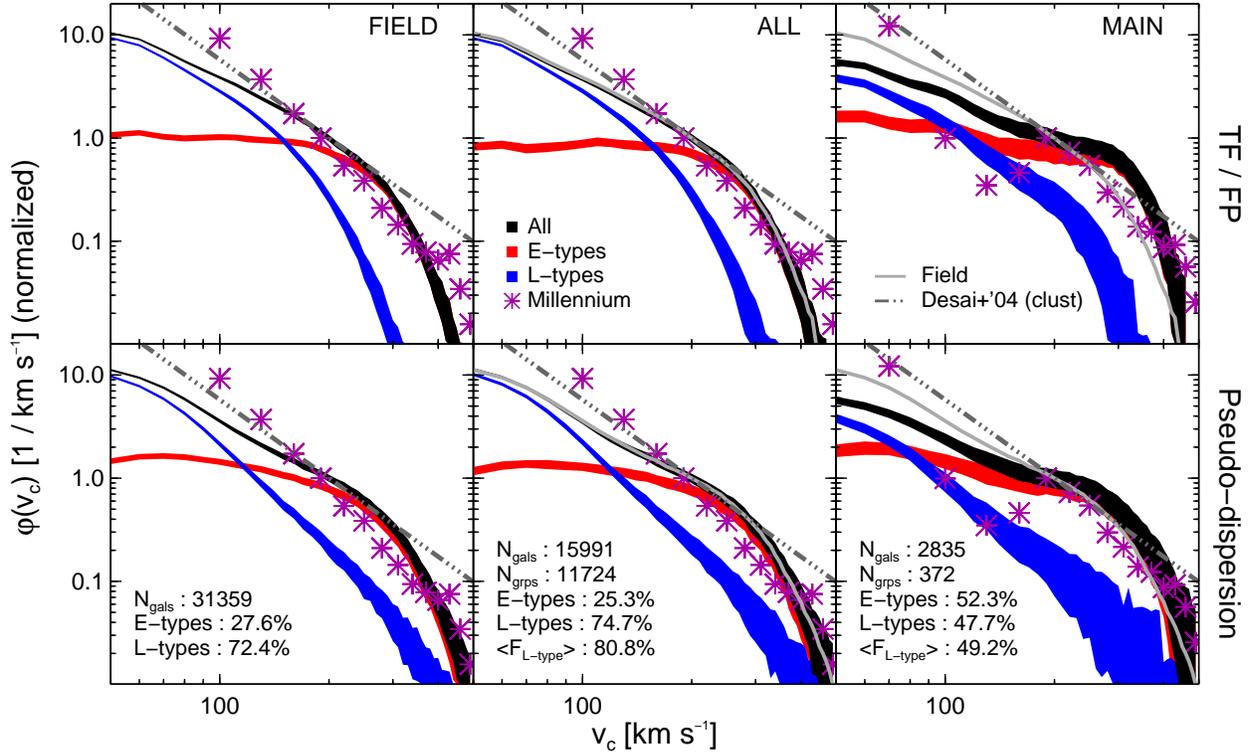}
\vskip -0.2cm
\caption{CVFs -- normalized to $\phi(200 \kms)$ -- of the three galaxy samples.  {\it Top}: TF/FP estimates.  {\it Bottom}: pseudo-dispersion estimates.  The ALL CVF is indistinguishable from that of the field { (replotted in light grey)}, but the low-\vc\ slope flattens significantly once low-mass, single-member ``groups" are culled  (MAIN sample). Charcoal dot-dashes show D04's best-fit composite cluster CVF.}
\label{fig:vc_comp}
\end{figure*}

Using photometric scaling relations avoids these problems.  Besides extending analyses to substantially lower \vc, photometric estimates are likely more accurate than fiber spectroscopy in the case of disk-dominated systems because they can be calibrated to gaseous emission at radii where $v_{\rm c,obs} \approx v_{\rm c, halo}$.  We therefore adopt these estimators in what follows, employing two quasi-independent methods.  We note in advance that, as shown in Section 4, both approaches yield similar results.

\subsection{Method 1: The Fundamental Plane and Tully-Fisher Relations}
\label{sec:TFFP}

Following a procedure similar to that of D04, we first compute \vc\ separately for early- and late-type galaxies using the ``inverse" Fundamental Plane \citep[FP --][]{Dressler87FP, DDFP87, Bernardi03} and Tully-Fisher relations \citep[TF --][]{Tully-Fisher77}, respectively, in $r$-band.  For the latter we adopt the scaling of \citet{Pizagno07} as adapted to $h = 1$ cosmology.  The calculation of \vc\ using the FP and TF relations is outlined in Appendices \ref{sec:AC} and \ref{sec:AD} (Equations \ref{eq:FPrel}, \ref{eq:TFrel}) \resp.  

We define ``early-types" to have $g-r \geq 0.20 - 0.03 M_r$ based on fitting the bimodality of the color-magnitude digram.   Our results are robust to choices in galaxy classification, but since each type is treated separately TF/FP CVFs formally depend on the adopted method.  For completeness, we explore this dependence in some detail in Section 5 and Appendix \ref{sec:AE} (see Figure \ref{fig:used_grp_cmd}).

\citet{Courteau97} and \citet{Mocz12} have also derived $r$-band TF relations.  Using these does not significantly change our results, but we choose the Pizagno et al.\ relation because it is based on longslit spectroscopy (unlike \cite{Mocz12}) and calibrated directly to SDSS photometry (unlike \citet{Courteau97}) over a magnitude range closer to that of our sample ($-22 < M_r < -18.5$).

\subsection{Method 2: Pseudo-dispersions}
\label{sec:pseudo}

As a cross-check, we also estimate \vc\ using ``pseudo-dispersions" \citep{TranPseudo04, Taylor10, Bezanson11}. This metric is designed to capture a galaxy's $R = \infty$ velocity dispersion, $\sigma_{\infty}$, based on its \sersic index, $n$, half-light radius, $R_{\rm e}$ (in kpc), and a stellar-to-dynamical mass mapping, $M_{{\rm d}, n}(\Mstel)$:
\begin{equation}
	\sigma_\infty = \sqrt{\frac{GM_{{\rm d},n}(\Mstel)}{K_{\rm v}(n) R_{\rm e}}} \equiv \frac{v_{\rm c}}{\sqrt{2}}.
\label{eq:siginf}
\end{equation}
The quantity:
\begin{equation}
	K_{\rm v}(n) \equiv \frac{73.32}{10.465 + (n - 0.95)^2} + 0.954,
\label{eq:Kv}
\end{equation}
is a virial prefix from \citet{Bertin02}.

Above, $n$ and $R_{\rm e}$ are drawn from the VAGC, but $M_{{\rm d},n}(\Mstel)$ must be supplied from elsewhere or derived. \citet{Taylor10} provide an empirical mapping, but it is valid only for $\log\Mstel/\Msun \gtrsim 10.5$. As our sample extends to much lower $\Mstel$, we derive a new relation.

This is accomplished by setting $\sigma_{\infty}$ to $\sigmav$ (measured from the SDSS spectra and listed in the VAGC), inverting Equation \ref{eq:siginf}, and fitting $\langle\log M_{{\rm d}, n}/\Msun\rangle$ in bins of 0.1 dex in stellar mass for $\log\Mstel/\Msun \geq 8.7$. Using the larger FIELD sample, we find:
\begin{eqnarray}
	 \langle \log M_{{\rm d},n} / \Msun \rangle &=& 30\pm2\nonumber\\ 
	 							         &-& (5.0\pm0.4)(\log\Mstel/\Msun) \\
								         &+& (0.31\pm0.02) (\log\Mstel/\Msun)^2\nonumber.
\end{eqnarray}
Substituting this mapping back into Equation \ref{eq:siginf} then effectively sets $\sigma_{\infty} = \langle\sigma_{\rm v}(\Mstel, n, R_{\rm e})\rangle \equiv v_{\rm c}/\sqrt{2}$.

Because they are measured in a $3\arcsec$ aperture, a concern is that it may be inappropriate to base dynamical masses (hence \vc) on SDSS velocity dispersions. However, the average offset between pseudo-dispersion- and TF-derived \vc\ is just $-14 \kms$. This drops to $-6 \kms$ if internal extinction corrections not captured in the pseudo-dispersion calibration are neglected (Appendix \ref{sec:AC}). Since TF is based on total galaxy luminosities and calibrated to direct, large-radius rotation measurements, this comparison suggests pseudo-dispersions are robust and provide a meaningful check of TF/FP results.

%

We see no significant environmental dependence in either this or the FP relation. \citet{Mocz12} find the same for the TF relation, thus all scalings are applied equally to field and group samples.  Estimator cross-comparisons are shown in Figure \ref{fig:metric_comp}.

\subsection{Error Estimation}

As shown in Figure \ref{fig:vc_err}, \vc\ uncertainties are taken as the scatter in this quantity at fixed $M_r$. This is calculated independently for each galaxy type and \vc\ metric except the TF  relation, where we adopt a constant scatter of 0.06 dex \citep[0.4 mag; see Table 5 of][]{Pizagno07}.

\subsection{The Efficacy of Scaling Relations}

The use of photometric estimators arguably reduces the advantages the CVF enjoys over the stellar mass function in terms of its relationship to theory.  This is clearly true in some sense: pseudo-dispersions are partially based on $\Mstel$.  However, { in another sense it is {\it not} true}.  Photometric estimators -- particularly the well-established TF and FP relations -- have been calibrated { and cross-checked} numerous times to directly observable quantities over a broad range in \vc\ and $\sigmav$.  The agreement we obtain internally between our two metrics (Figure \ref{fig:metric_comp}) and externally with \HI\ studies (Figure \ref{fig:intercomp}) gives us confidence that our results are robust.  

\begin{figure}[t!]
\hskip -0.2cm
\includegraphics[width = \columnwidth]{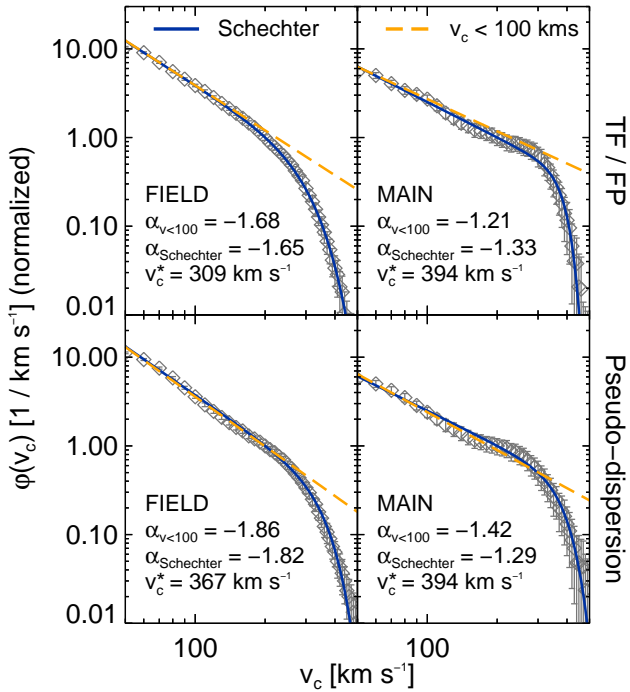}
\caption{Modified Schechter fits (solid blue curves; see Equation \ref{eq:schechter}) to FIELD (left) and MAIN (right) CVFs from Figure \ref{fig:vc_comp}. Grey diamonds are data; dashed orange lines are of the form $\log \phi \propto \alpha\log v_{\rm c}$ fit to $v_{\rm c} \leq 100 \kms$. See Table \ref{tbl:fit_params} for fit parameters.}
\label{fig:schechter}
\end{figure}


\section{Results}

\subsection{Group and Field Galaxy CVFs}

Figure \ref{fig:vc_comp} shows normalized CVFs, $\phi(v_{\rm c}) / \phi(200 \kms)$ for the three samples considered in this analysis.\footnote{The important discrepancy between field and cluster (or \lcdm) CVFs is their shape.  Normalizing to $v_{\rm c} = 200 \kms$ is arbitrary, but below the knee in the CVF.}  In columns 1, 2, and 3, \resp, these are: (1) the general field (FIELD); (2) all Y07 ``groups" (ALL); (3) all mass- and richness-complete groups ($\Nspec \geq 2$, $\log \Mgrp / \Msun \geq 11.0$; MAIN).  In the top row, we show results based on the TF/FP scaling relations; in the bottom, pseudo-dispersions.  The thickness of each band corresponds to $1\sigma$ uncertainties estimated by combining Poisson noise with the variance in each \vc\ bin across 100 Monte Carlo realizations with galaxies perturbed according to their error-bars.  In this and all similar plots, \vc\ binning is $10 \kms$ and red, blue, and black bands denote early-, late-type, and composite CVFs, \resp.

The right-most column reveals our main result: the CVF of the MAIN sample is significantly {\it flatter} at $v_{\rm c} \lesssim 200 \kms$ than that of the general field.  This suppression runs counter to the expected trend given the steeper CVFs found by D04 for cluster galaxies (charcoal dot-dashed line).  By $v_{\rm c} \approx 100 \kms$ the discrepancy between the CVFs is at least a factor of two.

This deficit is not due to biases in the Y07 sample. On a composite and type-by-type basis, the ALL and FIELD CVFs agree extremely well (middle column).  Late-type galaxies dominate the CVF at $v_{\rm c} \lesssim 150 \kms$, where the early-type CVF simultaneously flattens or turns-over \citep[see also][]{Sheth03, Choi07}.  Such agreement is unsurprising as the group catalog was constructed to reflect the full range of halo properties \citep{Yang07a} but it demonstrates that this goal is achieved in practice.

In all panels, we also plot the $z = 0$ CVF from the milli-Millennium simulation \citep[][]{Springel05}. All halos are included for FIELD and ALL sample comparisons, but only those within the appropriate $\Mhalo$ range ($12.4 \lesssim \log \Mhalo / \Msun \lesssim 14.3$) for MAIN group comparisons. {\it No further selection criteria are employed.} To construct the CVF, we simply bin all galaxy-harboring subhalos by their maximal {\it dark matter} rotation speed, {\tt vmax}. Subhalo occupation is determined by matching the galaxy catalog of \citet{DeLucia07} to the master ``MPAHalo" halo catalog.\footnote{see databases at \url{gavo.mpa-garching.mpg.de/Millennium/}} {\tt Vmax} is also obtained from the ``DeLucia" table, but we emphasize that it is a dark matter-based quantity. Thus, this CVF should be largely insensitive to the specific baryonic prescriptions these authors adopt. We have verified that results are unchanged if the models of \citet{Guo11} from the Millennium or Millennium-II simulations \citep{Boylan-Kolchin09} are used instead.

The power-law nature of the CVF from this dark-matter-only simulation is visible in the left two columns of Figure \ref{fig:vc_comp}.  However, the MAIN plots reveal this to be a super-position of separate distributions for central and satellite galaxies.  (We explore this further, below.)  Divergence between theoretical and observed CVFs is substantial at $v_{\rm c} \lesssim 150 \kms$, but agreement may be enhanced for late-types if an offset is applied.

To quantify the shape of our CVFs and aid future comparisons, we fit modified Schechter functions of the form: 
\begin{equation}
	\phi(v_{\rm c})dv_{\rm c} = \phi_{200}\left(\frac{v_{\rm c}}{v_{\rm c}^*}\right)^{\alpha}\exp\left[-\left(\frac{v_{\rm c}}{v_{\rm c}^*}\right)^{\beta}\right]\frac{dv_{\rm c}}{v_{\rm c}^*},
\label{eq:schechter}
\end{equation}
to those of the FIELD and MAIN samples.  Here, $v_{\rm c}^*$ is the ``knee" of the function, $\alpha$ and $\beta$ the low- and high-\vc\ slopes, \resp, and $\phi_{200}$ set such that $\phi(200 \kms) = 1$. Table \ref{tbl:fit_params} lists parameter values for the fits shown in Figure \ref{fig:schechter}.\footnote{$\beta$ comes from the fundamental mapping: $L\propto\Mstel\propto v_{\rm c}^\beta$; its inclusion is necessary to capture the high-\vc\ cut-off. More complete descriptions of the data may exist, but a single Shechter function characterizes the CVF sufficiently for our purposes.} 

Using either scaling relation, $\Delta\alpha \equiv \alpha_{\rm groups} - \alpha_{\rm field} \approx 0.3$ to 0.5, signifying the flattening in the median group CVF is real at the $5$--$10\sigma$ level.  (More-positive $\alpha$ implies a more-depressed CVF at low-\vc.)  Covariance between parameters is high, but fitting a power-law over the range $60 \kms \leq v_{\rm c} \leq 100 \kms$ yields the same trends and consistent $\Delta\alpha$.  All fits return $-1.9 \leq \alpha_{\rm field} \leq -1.6$, significantly flatter than the \lcdm\ prediction.

As a final test, we re-constructed CVFs after culling galaxies with $\log\Mstel / \Msun < 9$ -- approximately 0.3 dex above our adopted completeness limit -- to verify that mass-incompleteness does not affect our findings.  Trends are unchanged, so we consider the full MAIN sample robust for our analysis.

\begin{figure}[t!]
\hskip -1.2cm
\includegraphics[width = 1.1\columnwidth, trim = 0cm 0.5cm 0cm 0.2cm]{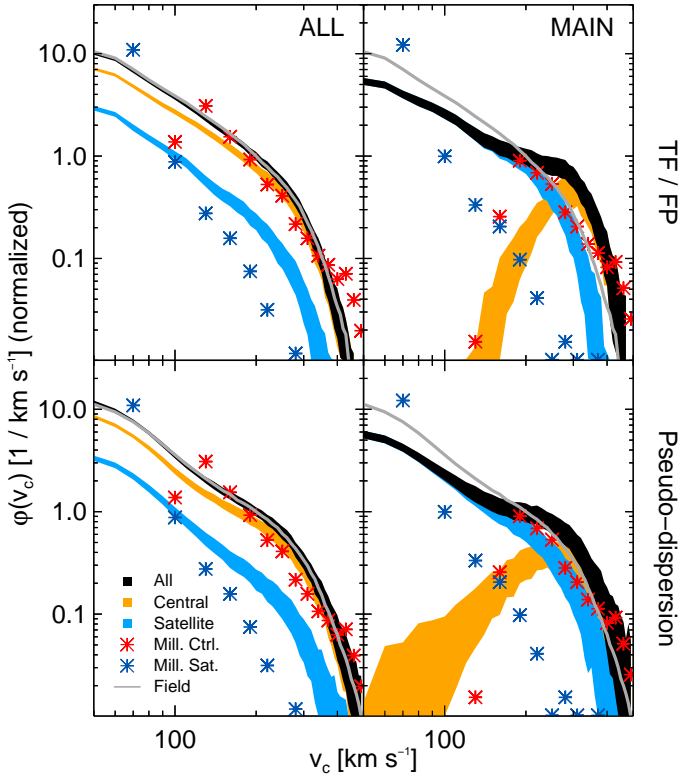}
\caption{Identical to columns 2 and 3 of Figure \ref{fig:vc_comp} but highlighting most-massive galaxies (``centrals") and other members (``satellites").  The ALL sample is dominated by low-mass centrals (i.e., isolated galaxies), whereas MAIN groups exhibit the central/satellite decoupling -- though at higher \vc\ -- displayed by Millennium halos of similar mass.}
\label{fig:vc_mmg}
\end{figure}

\subsection{Centrals and Satellites}

We re-plot our results to highlight a different aspect of the CVF in Figure \ref{fig:vc_mmg}.  Instead of early- and late-types, here we split ALL and MAIN groups into most-massive galaxies (``centrals") and everything else (``satellites").  

An interesting property of the ALL sample -- $\langle\Nspec\rangle = 1.3$ -- emerges from this perspective: its CVF is supported by centrals at {\it all} \vc.  This implies that low-mass, isolated galaxies\footnote{``Groups" with $\Nspec = 1$.} -- overwhelmingly late-types (Figure \ref{fig:vc_comp}, middle) -- dominate the low-\vc\ tail.  This result is key; we will return to it in Section 5.

In the MAIN sample -- $\langle \Nspec\rangle = 7.1$ -- a clear central-to-satellite transition is seen, taking place at $v_{\rm c} \sim 50$--70\% higher than that predicted by the Millennium simulation.

\subsection{Trends with Group Stellar Mass}

The composite CVF of MAIN groups thus appears shallower than both the field and cluster galaxy CVFs at $v_{\rm c} \lesssim 200 \kms$.  Does this divergence change as a function of group mass?

\begin{figure}[b!]
\centering
\includegraphics[width = \columnwidth, trim = 0cm 0.5cm 0cm 0.25cm]{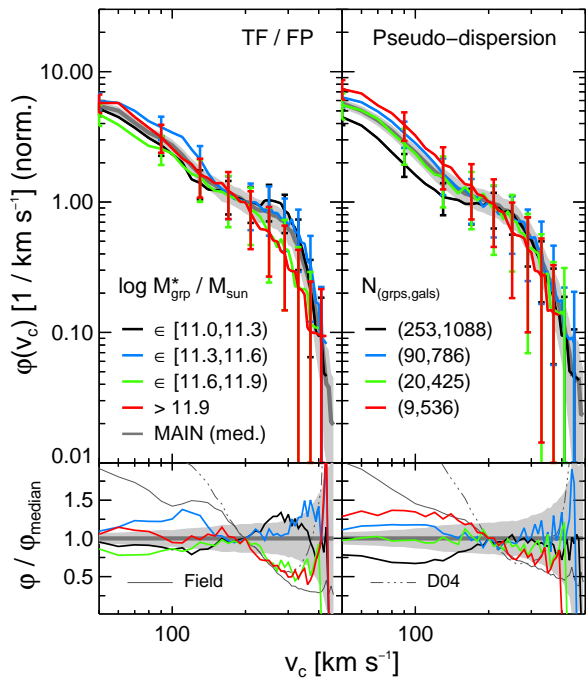}
\vskip -0.3cm
\caption{CVFs for MAIN groups in bins of $\Mgrp$ compared to the full sample.  There is no strong trend with increasing mass: all CVFs lie below the that of the field at low-\vc, though the magnitude of this suppression and the behavior at $v_{\rm c} > 200 \kms$ are mildly estimator-dependent.  A subset of error-bars is plotted.}
\label{fig:multi_mass}
\end{figure}

To explore this question, we split the MAIN groups into four bins of stellar mass and constructed the CVF for each sub-sample. Figure \ref{fig:multi_mass} presents these results.

\begin{figure*}[t!]
\centering
\hskip -0.2cm
\includegraphics[scale = 0.8, trim = 0cm 0.25cm 0cm 0.5cm]{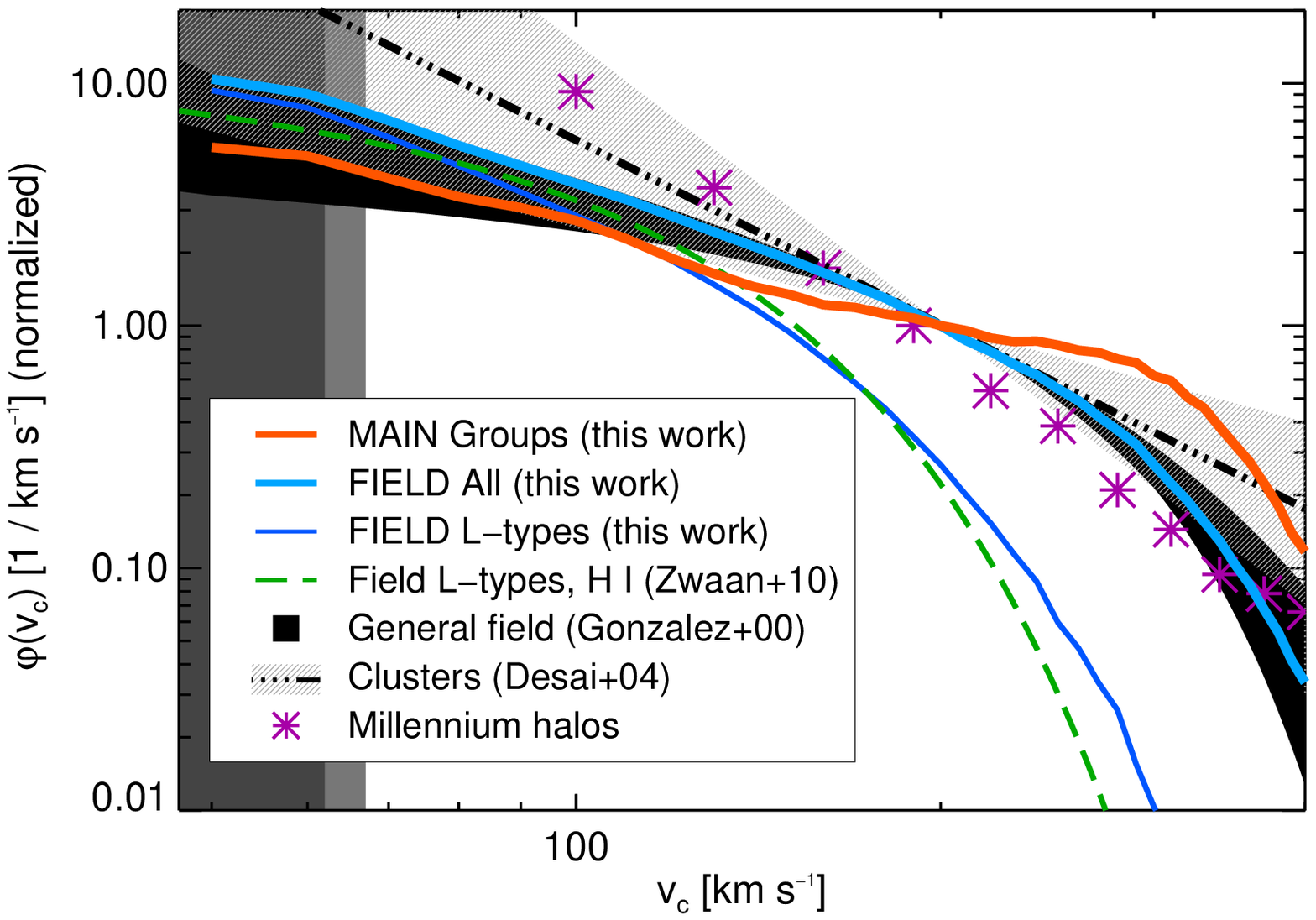}
\caption{Normalized CVFs in different environments from this and other works.  {\it Orange and turquoise curves}: our MAIN group and FIELD CVFs, \resp\  (TF/FP estimates).  {\it Grey hashes}: full spread of D04 cluster CVFs with best-fit composite power-law (black dot-dashes). {\it Black band}: full spread of field CVFs from \citet{Gonzalez00}. {\it Green dashes}: CVF from \HI\ data for {\it late-type} field galaxies from \citet[][complete to $30 \kms$]{Zwaan10}. {\it Thin blue curve}: our TF/FP CVF for field late-types, directly comparable to the Zwaan et al.\ CVF. Agreement between all field results -- especially our late-type CVF and the Zwaan et al.\ result --  is encouraging, reinforcing the accuracy of the TF scaling relation (which governs our CVFs at low-\vc) and our typing scheme. The MAIN group CVF is flatter than most field CVFs and all but the shallowest D04 cluster CVFs. Light/dark grey vertical regions show 75\%/50\% D04 completeness.}
\label{fig:intercomp}
\end{figure*}

At {\it all} $\Mgrp$, the CVFs appear similar; fluctuations are $\lesssim 50\%$ relative to the median distribution, often close to its uncertainties (bottom panel). There is mild divergence at the extremes of the mass range, with the CVF for highest-mass groups converging toward that of the field, but it is unclear if this is due to a deficit near $v_{\rm c}^*$ (TF/FP) or a steepening of $\alpha$ (pseudo-dispersion estimate).  A drop near $v_{\rm c}^*$ is expected since the central-to-satellite ratio is higher in low-mass groups, but regardless: such variations amount to perhaps a factor of two, strikingly small given the factor of ten separating these systems in mass and richness. Suppression relative to the field persists -- if marginally -- everywhere. Certainly, no {\it steepening} relative to the field is visible.

Notably, \citet{Calvi13} also find the $z\approx0$ group galaxy {\it stellar mass} function to be mass-independent and suppressed relative to that of isolated galaxies (if not the full field), qualitatively consistent with our results.

\subsection{Catalog Dependence}

To verify the above results, we repeated our analyses using the group catalog of \citet[][hereafter T11]{Tinker11}. The key difference between this and the Y07 sample is that T11 groups are selected from an {\it ab inito} galaxy mass-limited sample. Formally, this is different from drawing an effectively mass-limited sub-sample from a magnitude-limited catalog as we have done so far.

Due to the relatively high $\Mstel$ limit in the T11 sample -- $\log\Mstel /\Msun \geq 9.4$ -- we cannot probe as far down the CVF, but at $v_{\rm c} \gtrsim 120 \kms$ the picture does not change dramatically from that just presented.  We find all CVFs to be similar to that of the median result except for the same divergence -- here mostly near $v_{\rm c}^*$  using the TF/FP estimator --  at the extremes of the $\Mgrp$ range.  The most notable difference between the T11 and Y07 results is the disappearance of all significant low-\vc\ suppression compared to the field CVF (truncated to $\log\Mstel /\Msun \geq 9.4$), except when using psuedo-dispersions for groups with $\log\Mgrp/\Msun < 11.3$.

The unambiguous result from either sample is that the group galaxy CVF is {\it not} described by a power-law and varies by at most a factor of two at any \vc\ over more than an order of magnitude in $\Mgrp$.  Hence, the mass independence and {\it general} shallowness of the group galaxy CVF appear robust to group/estimator selection, though its suppression relative to the field CVF may be less so.

\subsection{Comparison with Previous Results}
\label{sec:comparison}

We seek to characterize the group galaxy CVF and any trends it might display with $\Mgrp$.  Self-consistent, controlled, relative analyses such as that described above are perhaps the best way to achieve this. However, comparing our results with previous measurements provides useful verification of their robustness and better-places field, cluster, and theoretical CVFs in context.

Towards this end, we set our TF/FP CVFs against the field CVFs of \citet{Gonzalez00} and \citet{Zwaan10} in Figure \ref{fig:intercomp}.  These studies are especially useful as they employed complementary methods to our own.  \citet{Gonzalez00} derived CVFs by transforming luminosity functions using the TF/FP relations while \citet{Zwaan10} used \HI\ data to measure it directly (though only for late-types).  All are normalized to their value at $v_{\rm c} = 200 \kms$ except the \citet{Zwaan10} CVF, which is normalized to the average of our late-type CVFs at that \vc. D04 cluster and Millennium Simulation results are overplotted for consistency.

Encouragingly, our field results coincide well with these measurements. Especially affirming is the extremely good agreement between our TF field late-type CVF and that of \citet{Zwaan10}, suggesting that our galaxy classification as well as \vc\ estimation process  for late-types  -- which control the low-\vc\ slope -- is robust (\HI\ is the most direct \vc-tracer known). The significance of the low-\vc\ deficits in our group CVFs is also better illustrated here, as is the general departure of results from \lcdm\ theory.  \citet{Pisano11} do not provide fits to their \HI-derived CVFs for Local Group analogs, but the low-\vc\ suppression we see qualitatively supports their findings.

\begin{figure}[t!]
\hskip -0.6cm
\includegraphics[width = 1.1\columnwidth, trim = 0cm 0.3cm 0cm 0cm]{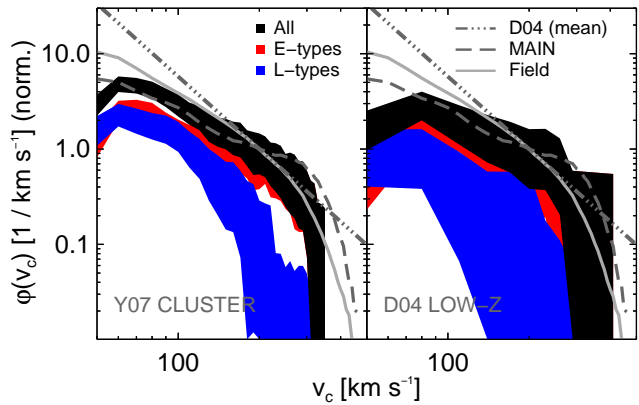}
\caption{{\it Left}: CVF (TF/FP estimate) of the largest Y07 group (Section 4.5).  {\it Right}: CVF of the two nearest clusters in the D04 sample.  Uncertainties are large, but no significant divergence is seen between either the two samples, or these clusters and the median MAIN group result (grey dashes).  The spectroscopic components of these systems all have $\Fltype \approx 0.3$, hence photometric extensions likely drive the steepening in $\alpha$ found by D04.}
\label{fig:clustocomp}
\end{figure}

Unfortunately, although the VAGC contains the D04 clusters, only two are close enough for us to study meaningfully ($z < 0.04$).  Desai et al.\ increased the depth of their sample by using the SDSS {\it photometric} catalog, statistically removing projected galaxies via a background subtraction technique. However, there is one cluster-sized halo in the Y07 catalog ($\log\Mhalo/\Msun = 15.1$, $\Nspec = 435$, $\zgrp = 0.0304$) for which we are complete to the mean D04 \vc\ limit ($\langle v_{\rm c, compl}\rangle \approx 62 \kms$).  Thus, besides yielding our own cluster measurement, comparing the CVF of this system to that of the {\it spectroscopic component} of the D04 clusters should reveal whether background { contamination} drives the discrepancies we see.  This comparison is shown in Figure \ref{fig:clustocomp}.  We use $R_{\rm vir}$ and $\sigma_{\rm cl}$ (cutting at twice this) from D04 (see their Table 1) to obtain spectroscopic cluster members from the VAGC.  
 
Examining this plot, we see that the Y07 (left) and two D04 clusters (right) display CVFs not only identical to each other (within admittedly large uncertainties), but also entirely compatible with the median MAIN group result.  For one of these systems, D04 quote a shallower slope than their mean result ($\alpha_{\rm cl} = -1.8$ and $\langle\alpha\rangle = -2.4$, \resp), but still comparable to that of our TF/FP field CVF ($\alpha = -1.7$) and steeper than what we find here.

The Y07 cluster is spectroscopically complete below the \vc\ limit of the (photometrically extended) D04 sample, so this comparison indeed suggests that photometric sources in the D04 analysis are responsible for the observed discrepancies in $\alpha$.

\begin{deluxetable}{lccrcc}
\tablecolumns{6}
\tablecaption{Modified Schechter Function Parameters}
\tablehead{
\colhead{Sample\tablenotemark{a}} &
\colhead{Estimate} &
\colhead{$v_{\rm c}^*$} &
\colhead{$\alpha$} &
\colhead{$\beta$} &
\colhead{$\phi_{200}$}\\
\colhead{} &
\colhead{} &
\colhead{($\rm km\ s^{-1}$)} &
\colhead{} &
\colhead{}
}
\startdata
Field & TF/FP	& $309\pm 8$ & $-1.65\pm 0.04$ & $3.4\pm 0.3$ & 188.5\\
Groups & TF/FP & $394\pm 7$ & $-1.33\pm 0.04$ & $9\pm 2$ & 160.4\\
Field & Pseudo & $367\pm 6$& $-1.82\pm 0.03$ & $4.8\pm 0.5$ & 128.2\\
Groups & Pseudo & $390\pm 11$ & $-1.29\pm 0.05$ & $6 \pm 1$ & 167.3
\enddata
\tablenotetext{a}{Field -- FIELD sample; Groups -- MAIN sample.}
\label{tbl:fit_params}
\end{deluxetable}

Finally, we note that the Y07 cluster is about 0.7 dex more massive than the largest systems in Figure \ref{fig:multi_mass}.  The agreement between its CVF and the MAIN median CVF thus reinforces the mass-independence discussed above. 


\section{Discussion}
\label{sec:discussion}

Our analysis yields three main findings: (1) the CVF of group galaxies is consistent with -- and possibly suppressed relative to -- that of the field; (2) its shape is grossly invariant across more than an order of magnitude of group stellar mass; (3) it is shallower than the cluster galaxy CVF of D04, even at comparable $\Mgrp$.  What drives these results?

\subsection{What Defines the CVF?}

Observationally, it is clear that the relative abundance of early- and late-type systems significantly alters the shape of the composite CVF (Figure \ref{fig:vc_comp}).  This is simply because the early-type CVF has flattened while the late-type CVF is still rising at velocities below $v_{\rm c} \sim 100 \kms$.  Hence, low-mass late-types substantially control the the low-\vc\ slope.

Because different scalings are applied to each class, CVFs estimated from the TF/FP relations depend explicitly on how ``early-" and ``late-types" are identified. Given the ``fuzziness" of galaxy classification \citep[e.g.,][]{Moresco13} this sensitivity may raise concerns.

We believe misclassifications are not a serious problem for two reasons, however.  First, identical trends emerge using the type-independent pseudo-dispersion estimator.  Second, results are qualitatively unchanged if any of three different categorization schemes are used (Appendix \ref{sec:AE}, Figure \ref{fig:used_grp_cmd}).  Indeed, because it is not a horribly inaccurate \vc\ estimator for {\it early-types} (Figure \ref{fig:vc_err}) tests show the basic shape of the CVF persists even if the TF relation is applied to {\it all} galaxies.  The same is true (at $v_{\rm c} \gtrsim 70 \kms$) if \vc\ is simply scaled from $\sigmav$ \citep[see also][]{Sheth03}.

Hence, astrophysically, the shape of the CVF appears to depend on the intrinsic mix of early- and late-type galaxies, not the way in which these terms are defined.

\begin{figure}[t!]
\hskip -0.4cm
\includegraphics[width = 1.1\columnwidth]{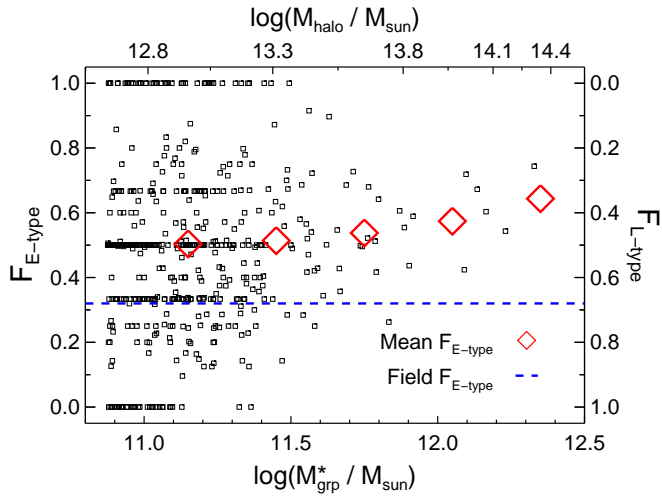}
\caption{Early-/late-type fractions for MAIN groups.  $\Fetype$ increases with $\Mhalo$, but the trend is weak and overwhelmed by scatter at $\log\Mhalo / \Msun  \lesssim 13.3$, contributing to the lack of a trend in CVF shape with group mass.  This figure is similar to Figure 2 of \citet[][]{Balogh10}.}
\label{fig:mean_etf}
\end{figure}

To demonstrate the effect of sample composition, we constructed CVFs for groups with late-type fractions higher than that of the general field, $\Fltype \geq 0.7$.  (Note that this is different than uniformly applying the TF relation to a mixed sample.)  These comprise about about 16\% and 78\% of the MAIN and ALL samples, \resp, highlighting again that the latter is composed mostly of isolated blue galaxies ($\Fltype \equiv 1$).  The MAIN sub-sample comprises all groups lying below the dashed line in Figure \ref{fig:mean_etf}. Figure \ref{fig:high_ltf} shows the CVFs.

The left column reveals, as anticipated, that the high-$\Fltype$ ALL CVF is steeper than that of the field.  However, the enhanced agreement between this CVF, that from D04 (below $200 \kms$), and the \lcdm\ prediction is unexpected. This suggests -- perhaps counterintuitively -- that dark-matter-only simulations best-describe isolated, gas-rich, star-forming galaxies, home to a wealth of baryonic processes not present in their ``red-and-dead" contemporaries. \citet{Blanton08} also found such agreement for a sample of isolated late-types and, as mentioned, \citet{Calvi13} saw a similar steepening in the mass function of isolated galaxies.  The overlap between this ``late-type enhanced" CVF and that from D04 -- built mainly from {\it early-type} galaxies in {\it dense} environments -- is therefore surprising.

The right column reveals hints that the high-$\Fltype$ MAIN group CVF may also steepen at low-\vc, but only using the pseudo-dispersion estimator. Conversely, though its low-\vc\ slope is consistent with that of the field or the median MAIN group CVF, the TF/FP CVF may agree better with \lcdm\ expectations by exhibiting a slight dip near the central/satellite transition predicted by the Millennium Simulation. 

Samples saturated with late-types may thus have CVFs whose shape is {\it roughly} consistent with theory or previous cluster results.  Unfortunately, they represent the galaxy population of neither the field nor groups nor clusters.

\subsection{Halo Mass Insensitivity and Group Make-up}

The general constancy in CVF shape we find as a function of group mass is seemingly at odds with the conclusions drawn above.  After all, if type fractions controlled its shape then the group galaxy CVF should na\"{i}vely converge to the early-type-only CVF with increasing $\Mhalo$ since their relative abundance ($\Fetype$) rises monotonically with this quantity \citep[e.g.,][]{Dressler80, PostmanGeller84, Dressler13}.

We do not expect to see such a trend for two reasons, however.  First, as also noted by \citet{Balogh10}, the variation in $\langle\Fetype\rangle$ from the lowest- to highest-mass groups in the MAIN sample is only $\sim 20\%$, much smaller than the scatter in this quantity at masses where most of our systems lie (Figure \ref{fig:mean_etf}).  Second, although $\Fetype$ rises, the central-to-satellite ratio drops with increasing $\Mgrp$, depressing the CVF at $v_{\rm c} \gtrsim v_{\rm c}^*$ and effectively steepening the low-\vc\ slope (Figure \ref{fig:multi_mass}).  The combination of these factors would subdue any low-\vc\ trends in the CVF.  

This is again in qualitative agreement with \citet{Calvi13}, who found the stellar mass function of group galaxies to be independent of parent halo mass.  Interestingly, although we disagree on its value, D04 also found the slope of the CVF to be independent of cluster velocity dispersion, a proxy for $\Mhalo$.  Thus, environment appears subsidiary to sample composition, influencing the CVF only insomuch as it correlates with type fractions.

\subsection{Group and Cluster Galaxy CVFs}

If late-type fraction drives the observed trends, no environment should have a steeper CVF at low-\vc\ than that of the isolated field.  As is well known and illustrated for our sample in Figure \ref{fig:mean_etf}, overdense regions overwhelmingly display {\it reduced} blue/late-type/spiral fractions relative to the field.  The paucity is especially pronounced for low-mass late-types (e.g., \citealt{Davis76, Blanton01}, cf. \citealt{Mobasher03, Zehavi11}) which, as just discussed, substantially control the low-\vc\ slope.  

\begin{figure}[b!]
\hskip -0.9cm
\includegraphics[width = 1.1\columnwidth, trim = 0cm 0.4cm 0cm 0.5 cm]{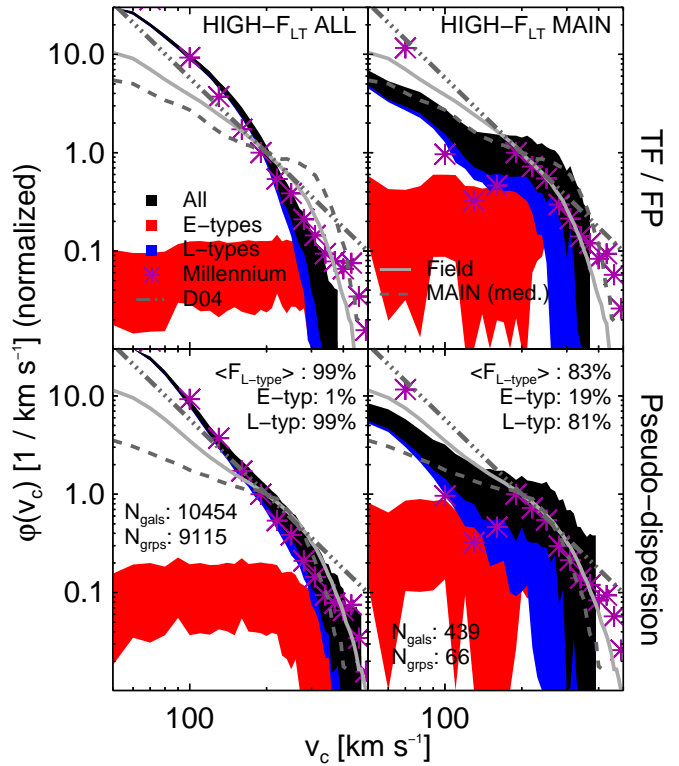}
\caption{CVFs from TF/FP (top) and pseudo-dispersion (bottom) scalings for ALL (left) and MAIN (right) groups with $\Fltype \geq 0.7$.  Especially in the ALL sample, the low-\vc\ slope steepens for these ``bluest" systems compared to the full catalog.  Further, this CVF is roughly consistent with the \lcdm\ prediction, as is the early-/late-type or central/satellite transition in the high-$\Fltype$ MAIN CVF.  Note that $\langle \Fltype\rangle \simeq 1$ for the former as it is dominated by isolated late-types.}
\label{fig:high_ltf}
\end{figure}

It is therefore unclear why D04 found the CVF of cluster galaxies to be most akin to that of the {\it bluest} sample we can construct and not our highest-mass groups.  The latter exhibit early-type fractions $\gtrsim 60$--80\%, similar to those found within cluster virial radii at $z \approx 0$ \citep[see again][and references therein]{Balogh10} and identical to that of the spectroscopic component of the two D04 systems we can study.

As our analysis of the one true Y07 cluster and this (albeit small) common sample yields a CVF consistent with the MAIN groups, and as both analyses employ almost identical \vc\ estimators, we can speculate only that the discrepancy is due to the incorporation of contaminated photometric sources by D04.  The steepening of the {\it luminosity} function for photometrically extended cluster samples has been discussed by \citet{Mobasher03} and seen in the simulations of \citet{Valotto01}, but repeat measurements using spectroscopic cluster samples should clarify the issue completely.

\subsection{Summary}

Motivated by  previous work on the field -- whose CVF is discrepant with theoretical predictions -- and clusters -- whose CVF was reported not to be -- we sought to determine if trends in the intermediate group environment revealed where this ``break-down" occurred, clarifying the utility of the CVF as an observational testing ground for galaxy formation in the \lcdm\ context.  Our results show that at all $\Mhalo$ the group galaxy CVF lies {\it at least as far} from model expectations as that of the field.

Of course, looming over this and all similar analyses is the assumption that \vc\ estimated from baryons corresponds to \vc\ predicted by (\Nbody) simulations.  Consensus on this point is not yet reached \citep[see e.g.,][]{MoMaoWhite98, Navarro00, Dutton12, Zolotov12} but at least \citet{Dutton10} suggest that, for late-types, the mapping should be close over most of the \vc\ range we probe.  Intriguingly, it is for these systems -- at least isolated ones -- that the best agreement with theory is seen both by us and \citet{Blanton08}.  This could suggest that merger processes -- likely involved in the formation of some early-type galaxies -- significantly decouple baryonic and dark matter dynamics, altering the relation between stellar $\sigmav$ and halo \vc\ in ways dark-matter-only $N$-body simulations do not capture.


Ultimately, given the results for isolated late-types, the CVF might {\it already} have demonstrated its virtue as a tool for testing \lcdm\ on galactic scales. However, one would like to extend the regime of reasonable comparisons substantially: many galaxies are neither isolated nor gas-rich.  The agreement of  results derived from photometric estimators and direct \HI\ measurements -- which probe very different radial scales, are sensitive to entirely different astrophysical influences, and have essentially no common systematic uncertainties --  suggests that further progress must be made in numerical modeling before the utility of the CVF as a proving-ground for theory can be fully assessed.


\section{Conclusions}

Using photometric estimators, we have constructed circular velocity functions (CVFs) for a diverse, mass-complete sample of $z \approx 0$ groups drawn form the SDSS.  Through self-consistent, controlled comparisons to field and cluster galaxies, we find:  
\begin{itemize} 
	\item The group galaxy CVF to be consistent with -- and possibly shallower than -- that of the field at $v_{\rm c} \lesssim 200 \kms$, in contrast to \lcdm\ predictions and previous cluster results.
	\item The shape of the CVF to be independent of halo mass up to $\log \Mhalo / \Msun = 15.1$.
	\item The shape of the CVF to depend mainly on sample composition, with increasing late-type fraction steepening the low-\vc\ slope.
	\item The CVF of isolated late-types to agree { better in shape} with dark-matter-only \lcdm\ predictions/simulations, which may also broadly capture the central/satellite transition in groups.
\end{itemize}

Future investigations using deeper spectroscopic catalogs \citep[e.g., the GAMA survey;][]{Robotham11} and upcoming IFU surveys such as SAMI \citep{Croom12}, MaNGA (\url{www.sdss3.org/future/manga.php}) and HETDEX \citep{Adams11, Hill11} may be able to shed significant light on the nature of these trends and discrepancies.


\section*{Acknowledgements}

We thank Dr.\ Jeremy Tinker for generously providing his group catalog, and Drs.\ Daniel Kelson, Alan Dressler, Ryan Quadri, Ned Taylor, Vandana Desai, Ann Zabludoff, Elena D'Onghia, and Daniel Masters for many helpful conversations. LEA acknowledges the website of David Fanning (\url{www.idlcoyote.com}) upon which he relied heavily. JSM acknowledges partial support for this work from NASA grant NNX12A161G.

Funding for the SDSS (\url{www.sdss.org}) has been provided by the Alfred P. Sloan Foundation, the Participating Institutions, the National Aeronautics and Space Administration, the National Science Foundation, the U.S. Department of Energy, the Japanese Monbukagakusho, and the Max Planck Society.

The SDSS is managed by the Astrophysical Research Consortium for the Participating Institutions. The Participating Institutions are The University of Chicago, Fermilab, the Institute for Advanced Study, the Japan Participation Group, The Johns Hopkins University, Los Alamos National Laboratory, the Max-Planck-Institute for Astronomy, the Max-Planck-Institute for Astrophysics, New Mexico State University, University of Pittsburgh, Princeton University, the United States Naval Observatory, and the University of Washington.\\


\bibliographystyle{apj}
\small\bibliography{lit.bib}


\appendix
\section{APPENDIX A: Incompleteness}\label{sec:AA}

As mentioned in Section \ref{sec:incompleteness}, both galaxy and group incompleteness are critical considerations when measuring the CVF.  

\begin{figure}[h!]
\centering
\includegraphics[width = 0.8\linewidth, trim = 0cm 0.5cm 0cm 0.5cm]{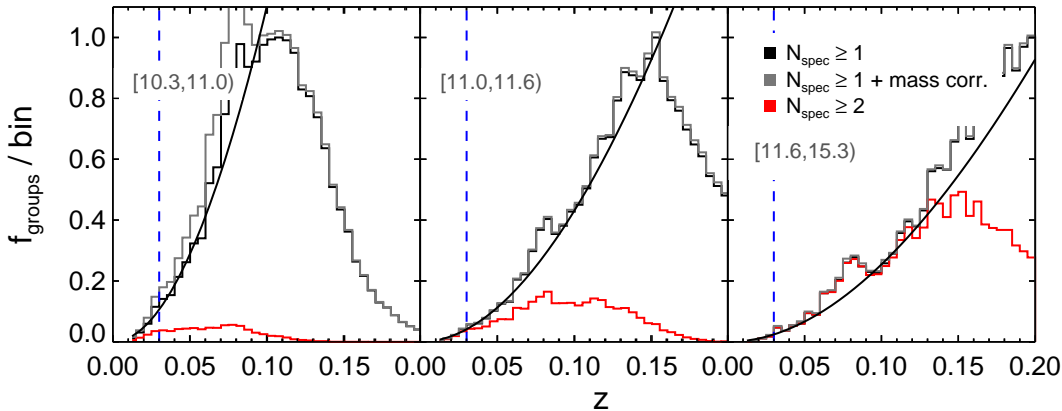}
\caption{Redshift distributions for groups in bins of $\log\Mgrp/\Msun$ (bracketed quantities).  {\it Black histograms}: groups with positive Y07 mass estimates (identical to bottom panel in Figure 6 of Y07). {\it Grey histograms}: the same selection after correcting for massless groups.  {\it Red histograms}: groups with $\Nspec \geq 2$.  Black curves trace the expectation for a homogeneous distribution of groups. The blue vertical dashed lines show the SDSS completeness limit for {\it galaxies} with $\log \Mstel / \Msun = 8.7$.  Samples with $\Nspec \geq 2$ become incomplete at substantially lower-$z$ than the full sample.  Of these systems only those with $\log \Mgrp / \Msun \geq 11.0$ are complete to the galaxy completeness limit.  Correcting for massless groups affects only the lowest group mass bin -- which we ignore -- and virtually no multiple-member systems.}
\label{fig:grpz}
\end{figure}

Regarding groups, various properties (e.g., $\Mhalo$) can be calibrated or corrected (e.g., for edge and fiber collision effects) using bootstrapping and mock catalogs (see Y07, \S\S 3.2--3.4).  However, a group catalog is ultimately based on a galaxy catalog and systems must drop-out at redshifts where enough of their members fall below the SDSS spectroscopic limit ($r \approx 18.0$) to prevent group identification.  Groups can also drop-out if their geometry or contents fail somehow to conform to halo identification criteria (see Y07, \S3).  We illustrate the impact of these effects in Figure \ref{fig:grpz}, showing the redshift distribution of groups in bins of $\Mgrp$.  Black histograms and rising curves -- identical to those from Y07 Figure 6 -- trace all groups ($\Nspec \geq 1$) and the expectation assuming uniform space density, \resp.  Grey histograms trace the same groups after correcting for null $\Mgrp$ entries (see Section \ref{sec:grpcat} and Appendix \ref{sec:AB}).  Red histograms trace groups with $\Nspec \geq 2$.  

Unsurprisingly, groups of different $\Mgrp$ become incomplete at different redshifts.  However, for groups with $\Nspec \geq 2$ there is no mass for which the catalog is complete to the sample redshift limit of $z = 0.2$.  Conversely, for groups of $\log \Mgrp / \Msun \leq 11.0$ ($\log \Mhalo / \Msun \lesssim 12.5$) there is no volume for which the group catalog could be considered ``complete" at all.  Hence, group mass and redshift cuts must be imposed before considering galaxy incompleteness.

{ Yet, for at least two reasons, the latter is the more significant problem.}  First, statistical corrections that assume a homogeneous source distribution -- such as $1/V_{\rm max}$ \citep[e.g.,][]{Felten77} -- cannot be employed; although galaxies of a given characteristic (e.g., stellar mass, $\Mstel$) may meet this criterion, {\it galaxies living in halos of a given characteristic} (e.g., stellar mass, $\Mgrp$) {\it may not.}  Hence, if, like us, one wishes to learn about the conditional CVF, $\phi(v_{\rm c}\, |\, \Mgrp, \Nspec, \dots)$, the aforementioned issues of group incompleteness prohibit (or at least dramatically complicate) using such methods.\footnote{They may be formally applicable in a group-complete volume, but see next paragraph.}

{ Secondly, there is simply no way to inject ``missing" galaxies into groups without {\it assuming} a luminosity, mass, or circular-velocity function. Hence,} cuts in richness translate directly into cuts in redshift: richer systems lie by construction at lower $\zgrp$ because (1) fainter galaxies enter the spectroscopic catalog, and (2) fiber collisions become less important.  { (Minimum SDSS fiber spacing -- $55''$ -- corresponds to $\sim 100$ kpc at $z = 0.1$, but just $\sim 30$ kpc at $z = 0.03$, where it excludes many fewer neighbors.)

If group incompleteness hinders the use of $1/V_{\rm max}$ weighting, galaxy incompleteness makes it dangerous. If rich groups are treated as if they could have been found anywhere in the survey, then the impact of their low-mass members will be artificially boosted; they are assumed to fill a much larger volume than they actually do.} We further discuss the effects of redshift--richness covariance in Appendix \ref{sec:AF} (see Figures \ref{fig:n_m_z}, \ref{fig:vmax}).

To remedy this problem one can either assume group composition is static and construct composite CVFs, drawing low- and high-\vc\ sources from different redshifts (see D04) or restrict samples to luminosity-, mass-, or \vc-complete volumes. We do not wish to assume a non-evolving group population \citep[see e.g.,][]{Williams12} and hence adopt the second course of action.

\begin{figure}[h!]
\centering
\includegraphics[width = 0.725 \linewidth, trim = 0cm 0cm 0cm 2cm]{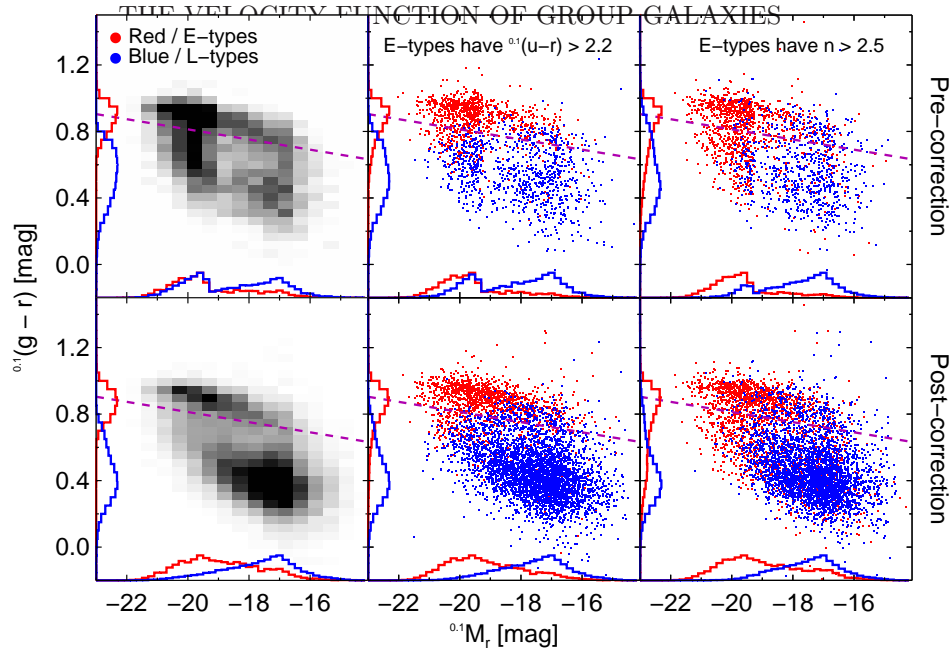}
\caption{Color-magnitude diagrams for the ALL sample using different galaxy classifications before (top) and after (bottom) correcting for ``massless" groups. From left, classification schemes are: our cut in $g-r$ versus $M_r$ (purple dashed line); the D04 cut in $u-r$ color; a cut in \sersic index, $n$.  The first two definitions are almost identical, the third tends to skew ``late-types" to fainter $M_r$.  For all definitions, ``missing" groups are preferentially composed of faint late-types. Failure to account for these groups does not bias samples of appropriately high $\Mgrp$ (see Figure \ref{fig:used_grp_cmd}) but can introduce artifacts in the CVF of the ALL sample (see Figure \ref{fig:nomasscorr_all}).}
\label{fig:all_grp_cmd}
\end{figure}

\section{APPENDIX B: The Effect of ``Massless" Groups}\label{sec:AB}

As mentioned in Section \ref{sec:grpcat}, many groups -- 58601 or $\sim 19\%$ of the Y07 catalog -- lack measured stellar or halo masses.  Na\"{\i}vely, one would simply exclude these systems.  However, because their contents is heavily biased, doing so can significantly distort global properties of the Y07 sample.

We illustrate two examples of this.  First, in the top { row} of Figure \ref{fig:all_grp_cmd} we show color-magnitude diagrams (CMDs) for ALL ``groups" with catalog entries for $\Mgrp$ and $\Mhalo$.  { The sharp density drop at $M_r \approx -19.5$ is due to} the Y07 group-finding algorithm: only galaxies brighter than this limit -- based on SDSS spectroscopic completeness at $z \leq 0.09$ -- are used to determine most group characteristics (e.g., $\Mgrp$, $\Mhalo$).

\begin{figure}[h!]
\centering
\includegraphics[width = 0.6\linewidth, trim = 0cm 0cm 0cm 1cm]{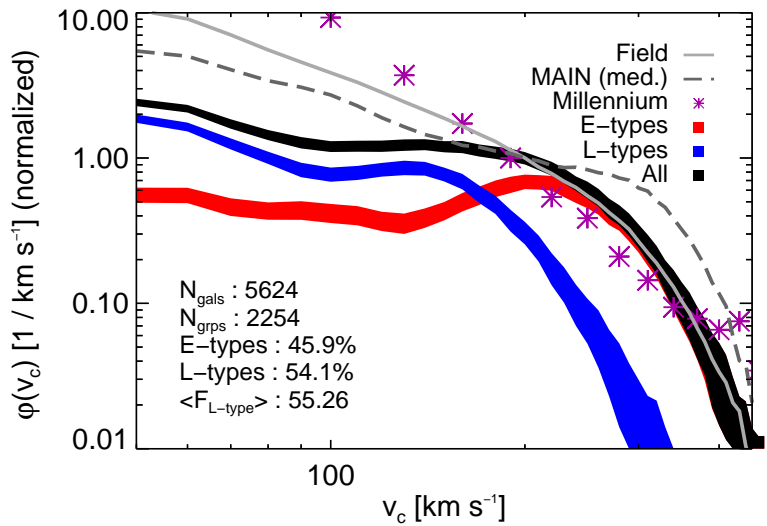}
\vskip -0.2cm
\caption{ALL CVF (TF/FP estimator) before { re-injecting} ``massless" groups.  Note the kinks in the distributions for both galaxy types at $v_{\rm c} \sim 100$--$130 \kms$ corresponding to the discontinuity visible in Figure \ref{fig:all_grp_cmd} (top).  Once all groups are assigned a mass the CVF for this sample converges to that of the field (see Figure \ref{fig:vc_comp}).}
\label{fig:nomasscorr_all}
\end{figure}

Of course, this discontinuity propagates to the CVF, which we show for the same ALL groups in Figure \ref{fig:nomasscorr_all}. The kink in the distributions for both types at $v_{\rm c} \sim 100$--$130 \kms$ reflects the same loss of galaxies with $M_r > -19.5$.  

We corrected for these effects by fitting a second-order polynomial to the relationship between total $\Mstel$ of member galaxies and $\Mgrp$ in systems for which the later quantity is known.  ``Massless" groups were assigned the $\Mgrp$ corresponding to the expected value given the mass of their members, which have $\Mstel$ estimates from the VAGC.  The distribution of these systems is traced by the grey histograms in Figure \ref{fig:grpz}.  From this and the two plots just discussed we can see that the ``missing" Y07 ``groups" are essentially all low-mass, isolated, late-type galaxies, perhaps the most important population in terms of the CVF!\footnote{Post-correction, the median ALL $\log\Mstel/\Msun$ drops from 9.8 to 8.9 while $\Fltype$ rises from 0.53 to 0.75.  FIELD values are 9.0 and 0.72, \resp.}  When these systems are re-injected into the sample, the CMDs fill-out (Figure \ref{fig:all_grp_cmd}, bottom row) and the CVF converges to that of the general field (Figure \ref{fig:vc_comp}, middle column).

{ Care should thus be taken by investigators seeking to characterize global properties of the Y07 catalog, but this issue is not important for the sample we are mostly concerned with} (MAIN; $\log\Mgrp/\Msun \geq 11,\ \Nspec \geq 2$).  Indeed, for $\Nspec \geq 2$ the pre- and post-correction distributions in Figure \ref{fig:grpz} are essentially identical (and the latter is not plotted).  For similar or more stringent cuts no analyses will be significantly biased by ignoring ``massless" groups.

\section{APPENDIX C: Computing $V_{\rm c}$ Using The Fundamental Plane}\label{sec:AC}

As originally defined, the Fundamental Plane \citep[FP --][]{DDFP87, Dressler87FP} related an elliptical galaxy's half-light radius, $R_{\rm e}$ (in kpc), to its velocity dispersion, $\sigmav$, and surface brightness, $I_0$.  However, as shown in D04, this relation can be ``inverted" to enable the estimation of $\sigmav$ from photometry alone.  The inverse FP is defined by: 
\begin{eqnarray}
	\log\sigmav &=& c_1 \log R_{\rm e} + c_2 \log I_0 + c_3 \equiv \log\left(\frac{v_{\rm c}}{\sqrt{2}}\right),
\label{eq:FPrel}
\end{eqnarray}
where $I_0$ is the average surface brightness within $R_{\rm e}$.  Following D04, we base $R_{\rm e}$ on a galaxy's apparent half-light radius, $R_0$ (in arcsec), and axis-ratio, $b/a$.  The former is taken from a single \sersic fit to a galaxy's 1D light profile and the latter from a 2D exponential fit.  The calculation is thus: $R_0 = R_{\rm fit}\sqrt{b/a}$.  We adopt VAGC $r$-band values for all quantities.

For a galaxy at redshift $z$ such that $k$-corrections are negligible:
\begin{equation}
	\log I_0 = -0.4 \left[r + 2.5 \log(2\pi R_0^2) - 10 \log(1 + z)\right].
\end{equation}

We re-determined the coefficients $(c_1,c_2,c_3) = (0.846,0.535,6.064)$ by minimizing:
\begin{equation}
	\chi^2 = \sum_{i=1}^{N_{\rm gals}}\left(\frac{\Delta}{\delta\sigmav}\right)^2_i,
\end{equation}
for {\it early-type galaxies, only,} over the range $2.0 \leq \log\sigmav \leq 2.6$ \citep{Bernardi03} where the quantity:
\begin{equation}
	\Delta_i^2 \equiv \frac{(\log\sigmav - c_1 \log R_{\rm e} - c_2 \log I_0 - c_3)_i^2}{1 + c_1^2 + c_2^2},
\end{equation}
is the $i$-th galaxy's perpendicular distance to the fit and $(\delta\sigmav)_i$ is the VAGC formal error in its spectroscopic velocity dispersion.

Note that $R_0$ enters both terms on the right-hand-side of Equation C1 and some scaling between $\sigmav$ and \vc\ -- customarily $v_{\rm c} = \sqrt{2}\sigmav$ (isothermal spherical halos) -- must be assumed.

\begin{figure}[h!]
\centering
\includegraphics[scale = 0.55]{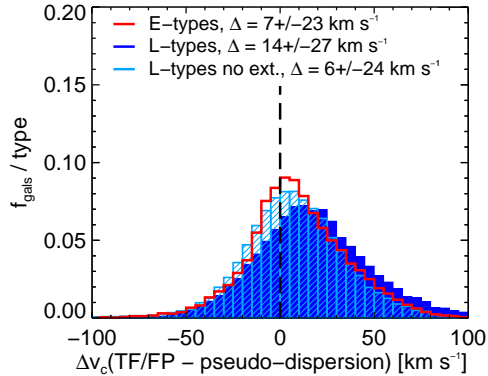}
\caption{Comparisons of \vc\ derived from TF/FP and pseudo-dispersion estimators. TF slightly overestimates \vc\ compared to pseudo-dispersions for late-types. Internal extinction corrections constitute $\sim 8 \kms$ of this offset.}
\label{fig:metric_comp}
\end{figure}

As illustrated in Figure \ref{fig:metric_comp}, estimates of \vc\ using the FP agree very well with those from pseudo-dispersions, exhibiting a mean offset of only $\langle\Delta v\rangle \equiv \langle v_{c,{\rm FP}} - v_{c, {\rm pseudo}}\rangle = 7 \kms$, and scatter $\sigma_{\Delta v} = 23 \kms$.

\section{APPENDIX D: Computing $V_{\rm c}$ Using The Tully-Fisher Relation}\label{sec:AD}

The TF relation \citep[][]{Tully-Fisher77} directly relates a spiral galaxy's circular velocity at $R = \infty$ (assumed to be the halo \vc) to its luminosity.  We break slightly from the procedure of D04\footnote{These authors use the $I$-band relation and scatter estimate of \citet{Giovanelli97}.  This may be marginally tighter than the $r$-band relation, but we wish to avoid large bandpass transformations (D04 take $I = M_r - 0.9$ from \citet{Fukugita95}).  Our results are qualitatively unchanged, however, if we construct CVFs using precisely their procedure.} and use the $r$-band relation of \citet[][hereafter P07]{Pizagno07} which is calibrated directly to SDSS photometry:
\begin{equation}
	\log v_{\rm c} \equiv \log v_{2.2} = 2.192 - \frac{M_r^{\rm corr} + [21.107 - 5\log( h / 0.7)] }{7.14}.
\label{eq:TFrel}
\end{equation}
Here, $M_r^{\rm corr}$ is $M_r$ corrected for inclination-dependent internal extinction using the formula of \citet{Tully98}:
\begin{eqnarray}
	M_r^{\rm corr}	&=& M_r - \gamma_r\log(a/b),\\
	\gamma_r &=& f(\gamma_B,\gamma_R, \lambda),\\
	\gamma_B &=& -0.35(15.6 + M_B + 5\log h_{80}),\\
	\gamma_R &=&  -0.24(16.2 + M_R + 5\log h_{80}),
\end{eqnarray}
where $a/b$ is the inverse $r$-band axis ratio, $h_{80} = H_0 / (80 \kms \Mpc^{-1}) = 0.8$ (here), and $\gamma_r$ is the $r$-band internal extinction, linearly-interpolated from $\gamma_B$ and $\gamma_R$.  Following P07 (see their \S4.4), this is done for each galaxy using the $gri \leftrightarrow BRI_{\rm C}$ bandpass transformations from \citet[][see their Table 7]{Smith02}, taking $\lambda_{(B,r,R)} = (438,617,641)$ nm.


Modulo inclination effects, the inverse TF relation is scatterless.  Hence, to construct late-type \vc\ error-bars we also adopt the P07 estimate of 0.061 dex ($\sim 15\%$) for its intrinsic width (see their Table 4).

We use the P07 relation for $v_{2.2}$ (\vc\ at 2.2 disk scale-lengths) as opposed to that for $v_{80}$ (\vc\ at the 80\% $i$-band light radius) as it has been shown to map closely to halo $v_{\rm c}$ by \citet{Dutton10} and to correspond well with \HI\ measurements by \citet{Courteau97}.  The relations are extremely similar, however, and no results are significantly affected if we use $v_{80}$ instead.

We have tested extensively the effects of using different TF relations.  Our results are qualitatively unchanged if the SDSS-derived TF relation of \citet[][based on {\it fiber} H$\alpha$ data]{Mocz12} or an inclination-averaged extinction correction \citep{Gonzalez00} are used.  Indeed, they are {\it quantitatively} unchanged if we adopt the relation of \citet[][from longslit H$\alpha$ data]{Courteau97} {\it and} an entirely different internal extinction prescription \citep{Tully85} {\it and} a \vc-dependent scatter estimate \citep[][see Eq. 11 therein]{Giovanelli97}.  { Combined with the very good agreement between it and the \HI-derived CVF for field late-types of \citet[][see Figure \ref{fig:intercomp} above]{Zwaan10}, these facts suggest} TF-derived CVFs are extremely robust.

{ Pseudo-dispersions underestimate TF-derived \vc\ by} $\langle\Delta v\rangle  = 14\kms$, within the scatter between the two metrics ($\sigma_{\Delta v} = 27 \kms$; see Figure \ref{fig:metric_comp}).  Internal extinction corrections account for $\sim 8 \kms$ of this offset.

\section{APPENDIX E: Galaxy Classification}\label{sec:AE}

As mentioned in Section \ref{sec:TFFP}, the detailed shape of composite CVFs derived from TF/FP estimates formally depends on the proportion of galaxies to which each relation is applied, as well as their intrinsic \vc\ distribution. { A worry is thus that the galaxy classification scheme might distort the CVF.}

We tested three classification methods:
\begin{enumerate}
	\item{A cut in the $g-r$ vs. $M_r$ plane; adopted here, early-types have $g-r \geq 0.20 - 0.03 M_r$ from a fit to the FIELD CMD.}
	\item{A cut in $u-r$ color; adopted by D04 from \citet{Strateva01}, early-types have $u-r \geq 2.22$.}
	\item{A cut in \sersic index, $n$, a proxy for a galaxy's structural/dynamical state; early-types have $n > 2.5$.}
\end{enumerate}

\begin{figure}[h!]
\centering
\includegraphics[width = 0.725\linewidth]{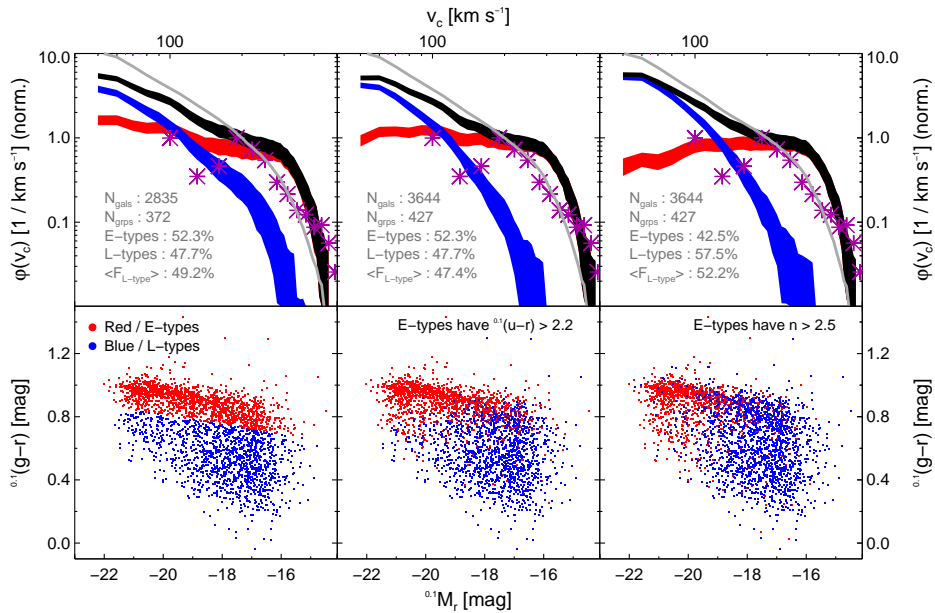}
\caption{CVFs and color-magnitude diagrams for various early-/late-type definitions for MAIN groups.  Column ordering is identical to that in Figure \ref{fig:all_grp_cmd}.  Results are qualitatively robust to which definition is used.}
\label{fig:used_grp_cmd}
\end{figure}

Figures \ref{fig:all_grp_cmd} and \ref{fig:used_grp_cmd} (bottom) illustrate these choices.  The left-most columns show the color-magnitude cut we employ, the middle columns the $u-r$ color cut, and the right the \sersic index cut.  As made plain by the top row of Figure \ref{fig:used_grp_cmd}, the MAIN CVF and its relationship to that of the field remain essentially unchanged regardless of the adopted definition.

We cannot identify a classification scheme that would produce a power-law CVF for group galaxies.  Even if all galaxies were categorized as ``late-types" and their \vc\ calculated using the TF relation the CVF would be Schechter-like.  However, as shown in Figure \ref{fig:high_ltf} and discussed in Section \ref{sec:discussion}, this does not imply that the mix of galaxy types has no effect on the CVF!\footnote{Indeed, samples saturated with isolated late-types -- insofar as these terms are meaningful -- exhibit a power-law-like CVF.}  It simply means that TF and FP \vc\ estimates for early-types are not wildly dissimilar and thus that our results are robust to large variations in galaxy typing.

Note that pseudo-dispersion-derived CVFs do not depend on galaxy classification since the metric is applied uniformly.

\section{APPENDIX F: Redshift--Richness Covariance}\label{sec:AF}

We initially supposed that the group galaxy CVF might depend on richness, $\Nspec$, as well as $\Mhalo$. We probed this dependence using the full Y07 sample ($0.01 \leq \zgrp \leq 0.20$), applying $1/V_{\rm max}$ weighting to all galaxies; large-scale structure, we believed, would wash-out over this volume.

Richness appeared to be {\it extremely} important using this approach: the richest systems at {\it all} group masses appeared to exhibit a common, power-law CVF, basically consistent with \lcdm\ predictions and D04 cluster results.  Further, the CVF was sensitive to specific richness values: at the highest $\Mgrp$, for example, CVFs for groups with  $\Nspec > 50$ appeared much more Schechter-like than those with $\Nspec > 100$.  The low-\vc\ slope flattened further as poorer groups were included.

\begin{figure}[h!]
\centering
\includegraphics[scale = 0.62, trim = 0cm 0cm 0cm 1cm]{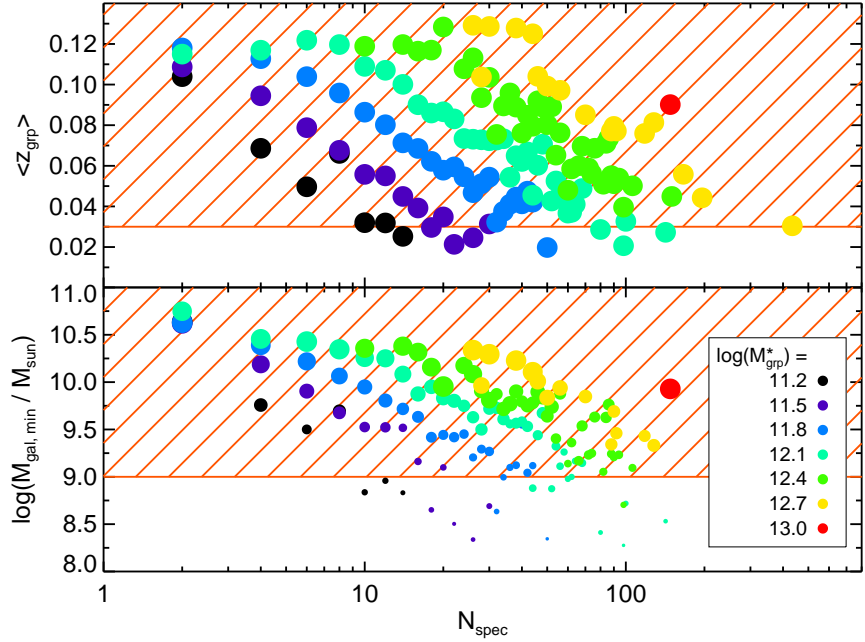}
\vskip -0.2cm
\caption{{\it Top}: average group redshift at a given richness in bins of $\Mgrp$.  {\it Bottom}: minimum $\Mstel$ probed in these groups; symbol size reflects $\langle \zgrp \rangle$ normalized maximum for each bin in the top panel.  Only groups with $\geq 2$ galaxies of $\log\Mstel/\Msun \geq 10.6$ are plotted to ensure minimum contamination.  At all group masses, cuts in richness translate almost one-to-one to cuts in redshift.  This fact complicates (if not prohibits) the use of galaxy incompleteness corrections based on the uniform density assumption; i.e., it biases analyses using $1/V_{\rm max}$ weighting (see Figure \ref{fig:vmax}).  Given our $\Mstel$ and richness criteria, our analysis was forced to exclude systems in the orange hatched regions.}
\label{fig:n_m_z}
\end{figure}

However, it became clear that these trends were spurious, driven by the $V_{\rm max}$ ``corrections".  The explanation is illustrated in Figure \ref{fig:n_m_z}.  

\begin{figure}[b!]
\centering
\includegraphics[scale = 0.62, trim = 0cm 0cm 0cm 1.75cm]{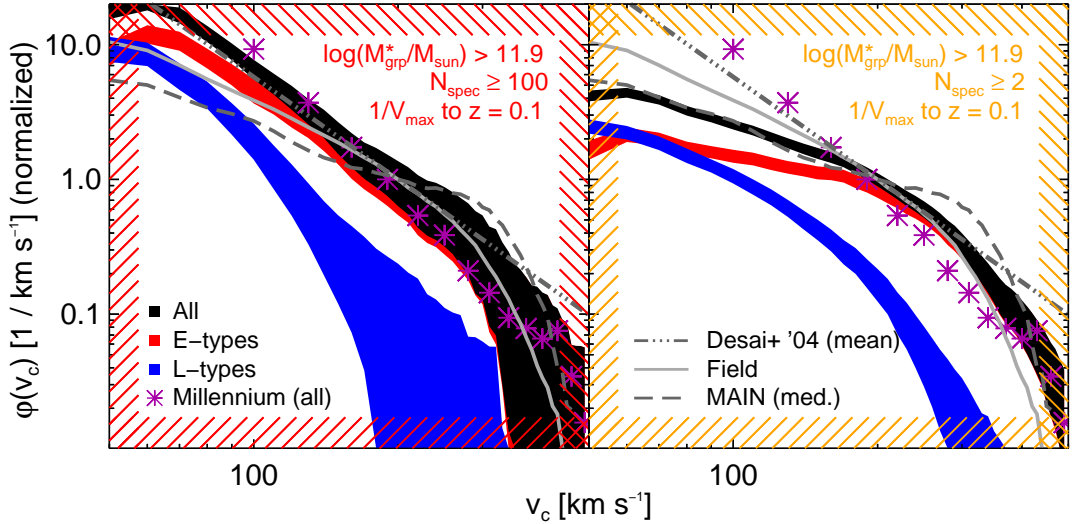}
\vskip -0.2cm
\caption{{\it Left}: CVF for high-mass ($\log \Mgrp / \Msun \geq 11.9$) rich ($\Nspec \geq 100$) groups with $\zgrp \leq 0.1$.  The richness cut is intended to select ``cluster-like" systems.  {\it Right}: CVF for groups of the same $\Mgrp$ but $\Nspec \geq 2$.  Galaxies have been weighted by $1/V_{\rm max}$ assuming uniform density to $z = 0.1$.  With this weighting, the ``cluster-like" CVF {\it appears} to agree extremely well with the full \lcdm\ prediction for all halos and the D04 results at $v_{\rm c} \lesssim 300 \kms$.  However, as illustrated in Figure \ref{fig:n_m_z}, for such rich groups {\it the uniform density assumption is false} ($N_{\rm grps} = 11$ with $z_{\rm median} = 0.05$, approximately 1/8 of the total volume) and the weighting  { inappropriate}.  Groups considered in the right panel much better meet this criterion, so the weighting is less biasing.  This CVF flattens significantly at low-\vc\ as also seen in similar mass systems in the volume-complete sample (see Figures \ref{fig:multi_mass} or \ref{fig:clustocomp}, left panels).}  
\label{fig:vmax}
\end{figure}

From this plot it is evident that richness is tightly anti-correlated with group redshift (top) and thus minimum galaxy $\Mstel$ probed (bottom).  This statement is equally true for groups of any $\Mgrp$.  Hence, selecting ever richer groups equates to selecting systems from ever decreasing volumes.  The power-law CVF shape emerges as members of these low-$z$ groups are inappropriately weighted {\it as if they filled the full survey volume}, boosting the low-\vc\ tail hugely compared to the high-\vc\ head (supported by galaxies with small $V_{\rm max}$ corrections).  The rather convenient effects of such ``up-weighting" -- empirically, $d\log(V/V_{\rm max})/d\log v_{\rm c} \approx -3.5$, coincidentally close to the theoretical slope of the CVF -- are illustrated in Figure \ref{fig:vmax}.

We discovered this problem when examining the CVF of the largest Y07 group, which fortuitously lies at $\zgrp = 0.0304$ (Section \ref{sec:comparison}).  Using the $V_{\rm max}$ method in the full survey volume ($\zgrp \leq 0.2$) galaxies in this group appeared to have a power-law CVF steeper than that of the field.  However, when truncating our survey to $\zgrp = 0.031$ the CVF of the same group became Schechter-like, suppressed compared to the field in good agreement with the median CVF of the MAIN sample (Figure \ref{fig:clustocomp}).

$\Mgrp$ and $\Nspec$ are well-correlated in the small volume considered above, so we did not pursue a richness-based analysis.

\clearpage

\end{document}